# Atomic Configuration of Nitrogen-Doped Single-Walled Carbon Nanotubes

Raul Arenal[1,2]*, Katia March [3], Chris P. Ewels[4], Xavier Rocquefelte[4], Mathieu Kociak [3], Annick Loiseau[5], Odile Stéphan[3]

[1]Laboratorio de Microscopias Avanzadas (LMA), Instituto de Nanociencia de Aragon (INA), Universidad de Zaragoza, Calle Mariano Esquillor, 50018 Zaragoza, Spain.

[2]ARAID Fundation, Calle Mariano de Luna, 50018 Zaragoza, Spain.

[3]Laboratoire de Physique des Solides (LPS), CNRS UMR 8502, Université Paris Sud XI, Bâtiment 510, 91405 Orsay, France.

[4] Institut des Matériaux Jean Rouxel (IMN), CNRS UMR6502, Université de Nantes, 44322 Nantes, France.

[5]Laboratoire d'Etude des Microstructures (LEM), UMR 104 CNRS-ONERA, 29 Avenue de la Division Leclerc, 92322 Châtillon, France.

*To whom correspondence should be addressed: E-mail: arenal@unizar.es

**Abstract**: Having access to the chemical environment at the atomic level of a dopant in a nanostructure is crucial for the understanding of its properties. We have performed atomically-resolved electron energy-loss spectroscopy to detect individual nitrogen dopants in single-walled carbon nanotubes and compared with first principles calculations. We demonstrate that nitrogen doping occurs as single atoms in different bonding configurations: graphitic-like and pyrrolic-like substitutional nitrogen neighbouring local lattice distortion such as Stone-Thrower-Wales defects. The stability under the electron beam of these nanotubes has been studied in two extreme cases of nitrogen incorporation content and configuration. These findings provide key information for the applications of these nanostructures.

Doped carbon nanotubes (NT), notably nitrogen-doped ($CN_x$-NT), have attracted much attention because of their interesting physical and chemical properties (*1-4*). Knowledge of the atomic arrangement of the dopant atoms in such nanostructures is essential for a complete understanding of the material's electronic properties, e.g. field emission (*5*) or transport (*6*) properties. This requires precision measurements, combining high spatial resolution and high spectroscopic sensitivity. Several techniques have been deployed with this aim, but until now none of them have provided the required information in such nanostructures. Most characterization techniques, such as Raman spectroscopy and x-ray absorption spectroscopy (XAS), have relatively low spatial resolution, of the order of hundreds of nanometers, and in some cases only provide indirect information (*1,2*). Thus, local structural and analytical methods are needed. Scanning tunneling microscopy (STM) and spectroscopy (STS) are local techniques for studying the structural and electronic features/properties of materials. Indeed, some studies have been reported on $CN_x$-NT (*7-9*) and recently also on N-doped graphene (*10,11*). However, chemical characterization cannot be unambiguously performed from the STM/STS analysis. Indeed, the interpretation of the results remains complicated, as different local structures may give rise to similar features (*7-11*). In this sense, high-resolution transmission electron microscopy (HRTEM) and scanning TEM (STEM) combined with electron energy-loss spectroscopy (EELS) have provided very valuable and rich information at the atomic scale (*12*).

Great progress has been made in (S)TEM due to the development of aberration–corrected microscopes (*12-14*). Recent work has demonstrated that annular dark-field imaging performed in a $C_s$-probe corrected STEM enables quantitative atom-by-atom analysis of 2D low-Z materials such as h-BN (*13*). Single atoms have also been investigated via STEM-EELS in these materials (*12, 14-21*). Three recent works have demonstrated the possibility to identify by STEM-EELS the presence of N atoms in graphene (*19-21*). In addition, N atoms have been also detected in N-implanted multi-walled carbon nanotubes, although not at the atomic-resolved scale (*22*). However, none of these studies have unambiguously provided, via EELS analysis, the local bonding configuration of these N atoms. Indeed, atomically-resolved EELS can yield not only the elemental composition, but also chemical bonding/local environmental information (*23,24*).

Single-walled nanotubes (SWNTs) pose significantly more complications for high resolution imagery and spectroscopy than graphene due to their high defect concentration and motional instability under the electron beam. Indeed, a clear and unambiguous proof of identification of individual dopants (N atoms in this case) in single-walled nanotubes has not yet been obtained, in contrast to the case of 2D materials (*25*), and the question remains whether electron microscopy is able to identify individual dopant atom configurations in such complex non-planar and beam-sensitive architectures. Here, we have carried out such experiments on single-walled, nitrogen-containing carbon nanotubes ($CN_x$-SWNTs) synthesized via a laser vaporization technique (*8,26*). We clearly demonstrate the feasibility of dopant identification in a nanotube at the single-atom level. Notably, we show that as well as discriminating between two individual atoms of low atomic number differing only by one electron (carbon Z=6 and nitrogen Z=7), this approach can also unambiguously identify the local chemical bonding of the atoms in topologically complex materials.

The HRTEM micrographs in Figure 1 demonstrate the presence of defects and unusual polygonal ends in the SWNTs. These images confirm previous studies suggesting the incorporation of nitrogen atoms in the honeycomb carbon structure of a NT, which usually implies the presence of defects (*1-3*). This is corroborated by the higher sensitivity of $CN_x$-SWNTs under the electron beam (see *Supplementary* Movies 1 and 2). Therefore, atomic-level (S)TEM studies on $CN_x$-SWNTs require low voltages to avoid knock-on damage (*27,28*). They also require mechanically stable NTs, with two fixed extremities to reduce vibrations, near to which the measurements can be taken (see *Supplementary* Fig. S1). These important aspects of the NTs behaviour under the electron beam limit the information that can be extracted from the

HRTEM images. Thus, the only way to achieve these goals is via analytical STEM-EELS measurements.

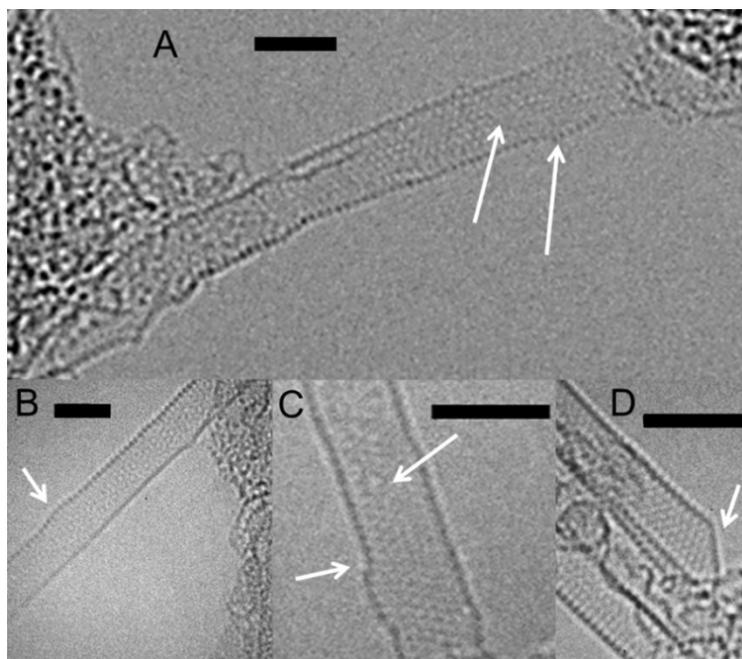

**Figure 1.** *Nitrogen doped single-walled carbon nanotubes: High-resolution transmission electron micrographs of different single-walled nitrogen containing carbon nanotubes ($CN_x$-SWNTs), taken at 80 kV. The images show the presence of defects within the walls marked by white arrows: side and central section of the imaged NTs. These latest defects correspond to vacancies, which may be associated to the presence of nitrogen atoms. Fig. 1(D) displays a flat end of a NT, a feature rarely observed in pure carbon single-walled NTs. The scale bar is 2 nm for all the micrographs.*

We have performed atomically-resolved EELS experiments at low-voltage (60 kV) and in high-vacuum conditions (~1.4 x $10^{-8}$ Torr) to avoid carbon contamination, comparing the resulting spectra to density functional simulations (details in *Supplementary* Information).

A high-angle annular dark field (HAADF) STEM image of an area containing three $CN_x$-SWNTs forming a bundle is shown in Figure 2A (associated bright field image in *Supplementary* Fig. S2). Figure 2B shows a zoom after filtering of the green rectangular region, where the atomic resolution of the lattice is clearly visible. One atom, labelled with an arrow, shows higher intensity than its neighbours (~1.2 times more intense), see profile intensity plotted in Fig. 2C. This intensity is consistent with the $Z^{1.7}$ contrast expected between C and N (1.15 assuming a superposition of the nitrogen atom with an underlying carbon).

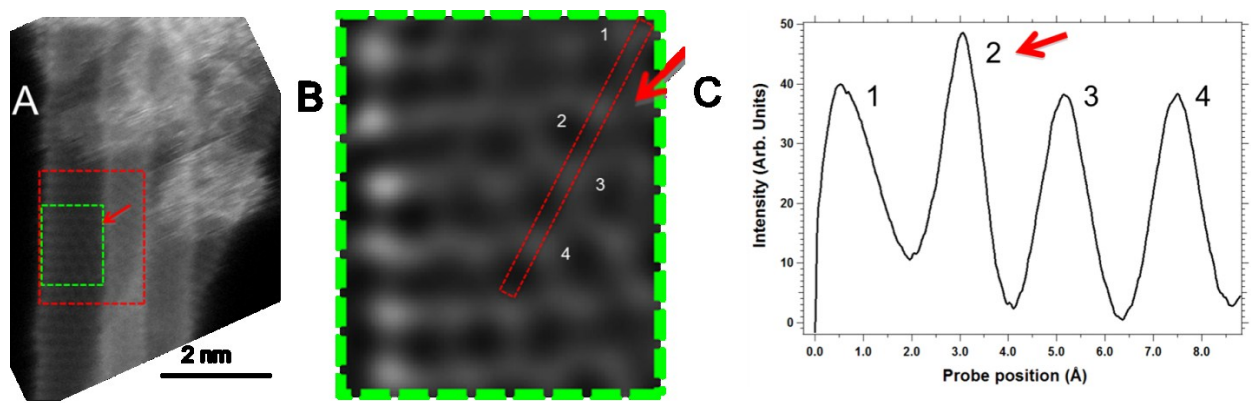

**Figure 2.** *HRTEM identification of a single substitutional nitrogen atom (A) HAADF image of a bundle of SWNTs. The red dotted rectangle indicates the position where the SI of Figure 3 has been recorded. (B) filtered image of the green dotted region in (A). 4 atoms are labelled from 1 to 4. The red arrow points toward brighter atom 2. (C) Intensity profile at the location of the red rectangle from Fig. 2B. The positions of atoms 1-4 are indicated. Brighter atom 2 is pointed by a red arrow.*

An EELS spectrum-image (SI) (*29,30*) was recorded over the area of Fig. 3A (marked in Figure 2A with a red dotted box), and the corresponding HAADF image captured simultaneously is shown in Figure 3B. Despite the coarse sampling (the pixel size was set to 0.5 Å so as to minimize the electron dose) the atomic structure of the nanotube is once again clearly visible. Ripples along the nanotube are evident during acquisition, reflecting its vibration under the electron beam. Fig. 3D displays three single EEL spectra extracted from this SI in the marked positions/pixels of Fig. 3B (spectra labelled (i), (ii) and (iii)). The carbon K edge is visible in the three spectra. In two of these spectra the nitrogen signal is also detectable, see Fig. 3D ((i) and (ii)). Nitrogen was detected in only these two spectra out of the whole dataset (1755 spectra). Their position corresponds to the higher intensity atom identified in the HAADF above, confirming that it is indeed a nitrogen atom as suggested from the analysis of Fig. 2. The only two individual spectra in which nitrogen is detected are located in diagonally adjacent pixels. Considering the EELS signal delocalization, one would expect an effective spatial extension of the N1s EELS signal over a 2×2 grid of 4 pixels of the SI (pixel size is 0.5 Å). However due to SWCNT vibration under the electron beam, only part of the atom signals are detected. Considering the dimensions of the NT, this means the nitrogen concentration in the probed area can be estimated at ~ 0.25 atom %.

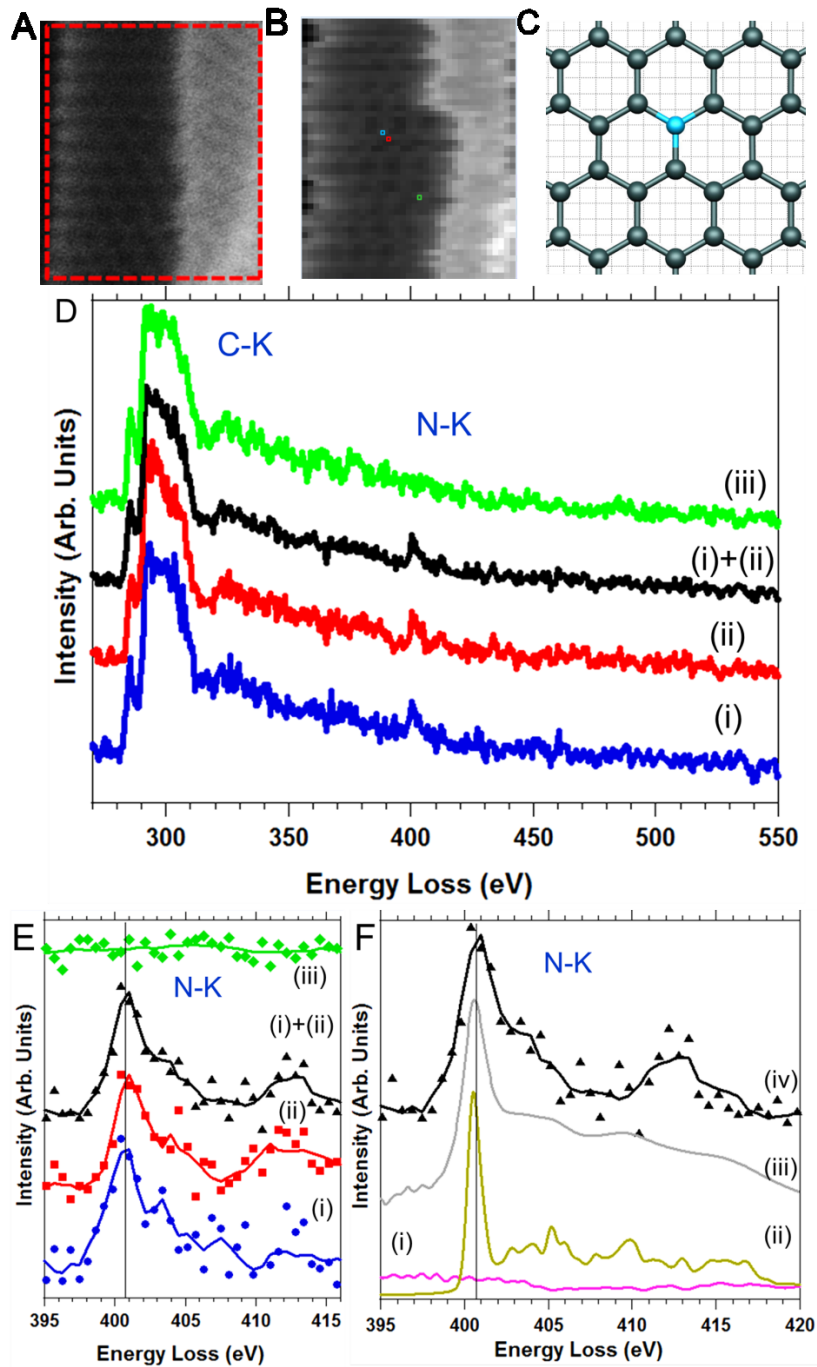

**Figure 3.** *Atomically-resolved EELS identification of a single substitutional nitrogen (A) HAADF image displaying atomic resolution. The red dotted rectangle marks out the scanned area during the acquisition of an EELS spectrum-image (39×45 recorded spectra (1.95×2.25 nm), step size 0.05 nm, probe size ~0.11 nm, acquisition time 100 ms/spectrum). (B) HAADF image acquired simultaneously with the spectra, showing the motional drift due to nanotube vibration. The only two pixels with N1s edge signal are outlined in blue and red. (C) DFT optimized structure of a substitutional nitrogen atom. Carbon and nitrogen atoms are respectively in gray and blue. (D) Selection of EEL spectra extracted from the SI; blue, red and green spectra correspond to the pixels outlined in blue, red and green in Fig. 3B. Each curve*

*corresponds to a single spectrum from the SI, except the black, which is a sum of the blue and red spectra. **(E)** N1s edge fine structures (ELNES), extracted from the spectra of Fig. 3D. In order to filter noise, spectra were smoothed using a Savitzky–Golay filter (second-order polynomial). **(F)** Simulated N1s ELNES (grey, (iii)) and nitrogen partial DOS calculations ((i) purple=$p_z$ $\pi^*$-states, (ii) green=$p_{x-y}$ $\sigma^*$ states) for substitutional nitrogen, compared to the experimental spectrum (iv), same as spectra (i)+(ii) of Fig. 3 E. The EELS simulations allow unambiguous assignment of the peak at ~401 eV in the N1s edge to a substitutional configuration shown in (C).*

The carbon 1s (C1s) energy loss near edge structure (ELNES) in Figure 3D show the peaks at ~ 285.5 eV and ~ 292 eV characteristic of $sp^2$-carbon materials, attributed to transitions from an initial 1s state to unoccupied anti-bonding-like $\pi^*$ and $\sigma^*$ states above the Fermi level (*2,14,17,18*). The nitrogen 1s (N1s) ELNES, expanded in Figure 3E, show a strong peak at ~401 eV, with very little signal at energies above this. Comparing the spectra to density functional theory (DFT) ELNES calculations of possible single nitrogen defects, there is excellent agreement with the spectrum for substitutional nitrogen (Fig. 3F(iii) and atomic model, Fig. 3C) across the range of $\pi^*$ and $\sigma^*$ bands. This assignment is consistent with the HAADF observations (Figures 2A, 2B) and with recent works on nitrogen doped graphene (*19-21*). The spectrum shows a number of signature features, which would be lost in a signal averaged with other nitrogen configurations. The $\pi^*$ peak is highly suppressed. This can be understood from the projected density of *p*-states (p-DOS) on the nitrogen atom (Figs. 3F(i) and 3F(ii)) and surrounding carbon neighbours. The additional nitrogen electron rests relatively localized and disrupts the local conjugation, reducing the local $\pi$- and $\pi^*$ contributions and resulting instead in a strong $p_z$-signal at the Fermi level, which is, not sampled (*Supplementary* Fig. S7). Instead there is an exceptionally sharp "molecular like" signature at the start of the $\sigma^*$ band, as shown in the partial $p_{x-y}$-DOS; the rest of the $\sigma^*$ signal is significantly reduced compared to bulk N1s edge EELS measurements for nitrogen doped nanotube samples in the literature *(26,31)*. We note that no apparent damage was observed after the acquisition of this SI (*Supplementary* Fig. S2). This is consistent with earlier calculations of electron beam knock-on cross-sections (*27*) suggesting that substitutional nitrogen and its neighbours should be stable under a 60 kV beam. This suggests that this configuration of substitutional graphitic nitrogen is a minority species in these bulk measurements carried out at higher acceleration voltages (100 kV instead 60 kV) and beam doses, with the probed defects demonstrating a much stronger $\sigma^*$ band at higher energies *(26,31)*.

To summarize thus far, this result demonstrates the capability of STEM EELS to successfully detect individual nitrogen heteroatoms within the wall of a SWCNT, as confirmed by HAADF imaging, as well as unambiguously identify the associated bonding configuration in combination with DFT ELNES simulations (in this case, substitutional "graphitic" nitrogen).

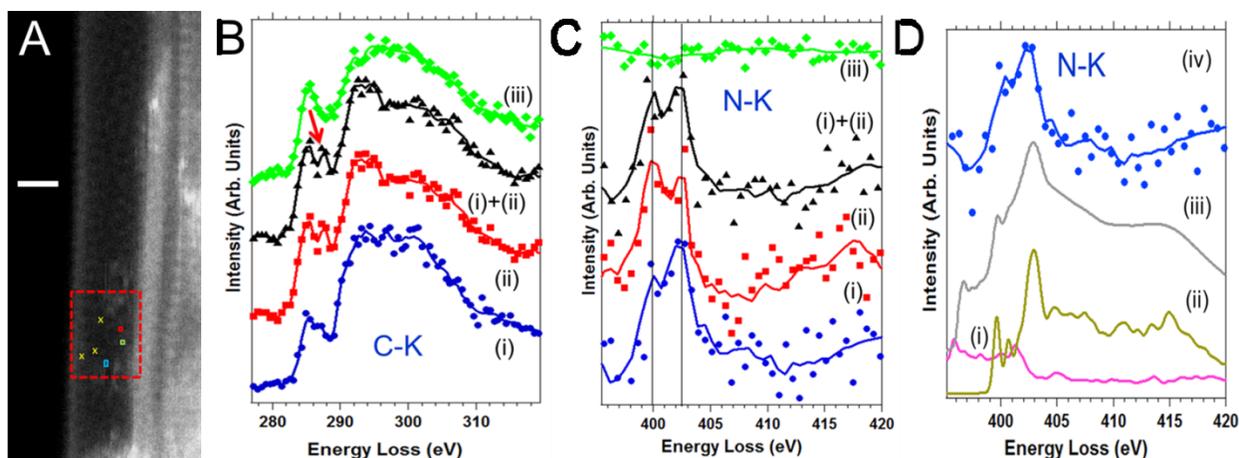

*Figure 4 Atomically-resolved EELS identification of "pyrrolic" nitrogen. (A) HAADF image obtained from a second bundle of single-walled nanotubes. An EELS SI was recorded from the rectangular area marked in the HAADF image (30x41 recorded spectra (1.5 x 2.05 nm), step size 0.05 nm with a probe size of ~ 0.11 nm, acquisition time 100 ms/spectrum). The scale bar is 1 nm. A selection of EEL spectra from the pixels marked on Fig. 4A (red, green and the 2 pixels in blue) are displayed in Supp. Fig. S3 (each spectrum is associated to a single pixel, except for the black one, which corresponds to the sum of the red and blue spectra). Yellow crosses indicate extra locations where nitrogen was detected but spectra not displayed. (B) C1s edge fine structures extracted from the spectra. A prominent peak at ~288 eV is observed in the spectra containing nitrogen (particularly in the red spectrum). (C) N1s ELNES obtained from each atomic position already mentioned. The N1s edge is observed in all these spectra, except in the green one which has been selected as a reference. Peaks at ~399.8 eV and ~402.7 eV are observed in two of these spectra. In order to filter noise, spectra were smoothed using a Savitzky–Golay filter (second-order polynomial). (D) Simulated N1s ELNES (grey, (iii)) and nitrogen partial DOS calculations ((i) purple=$p_z$ $\pi^*$-states, (ii) green=$p_{x-y}$ $\sigma^*$ states) for structure displayed in Figure 5(C), compared to the experimental spectrum (iv) from 4C(i).*

A more complex example is shown in Fig. 4. Fig. 4A shows an annular dark field image of a SWNT at the edge of the bundle (the corresponding bright field image is shown in *Supplementary* Fig. S4). A SI was collected from the rectangular area marked in Fig. 4A. Three different EEL spectra extracted from this SI are shown in Fig. S3, over the energy range covering the C1s, N1s and O1s edges. In contrast with the previous example (Fig. 2-3), nitrogen has been detected in 5 different positions, relatively far (a few angstroms) from each other, giving an estimated nitrogen content of ~ 1.6 atom %. For clarity, due to the low signal-to-noise ratio in some cases, we have focused the analysis on two of the positions for which the N1s edge is better resolved (spectra (i) and (ii); we also show the sum of these 2 spectra). It is worth mentioning that, during spectral acquisition, even though we worked under so-called "gentle-STEM conditions" (*28*), obvious beam-induced damage was observed after measurement (*Supplementary* Fig. S4 acquired after the SI). We have observed the difference in the behaviour of these two situations of $CN_x$-SWNTs, low and high structural instability, *via* HRTEM (*Supplementary* Movies 1 and 2), suggesting higher nitrogen concentration and associated intrinsic defect concentration. The C1s ELNES depicted in Fig. 4B show the characteristic profile of $sp^2$ carbon atoms in a hexagonal structure, with π* and σ* bands, as described above. This is the case for almost all the spectra in the SI. However, for most of the positions where nitrogen has been detected, there is an additional peak at ~288 eV, see Fig. 4B ((ii) and (i)+(ii)). This particular signature has previously been associated with several atomic configurations: C=O, C=C (from $CH_2$ groups, after reduction by the electron beam) and C-N bonds *(32)*. Its origin is revealed via analysis of the DFT ELNES simulations of the C1s edge (Fig. 5) for the substitutional, nitrogen-vacancy, and Stone-Thrower-Wales defects (two carbon atoms rotated about their bond centre, creating a patch of paired pentagons and heptagons in the lattice, see Fig. 5C). In the case of substitutional nitrogen the C1s edge of the surrounding atoms shows no significant deviation from those of carbon atoms in the pristine lattice far from the defect (see Figs. 5A and 5C). This is consistent with Figure 3D, where we see no C1s peak at 288 eV associated with substitutional nitrogen. However, in the case of the nitrogen-vacancy defect we see major variations, including a strong peak between the pristine π* and σ* bands, corresponding to our experimental observation at 288 eV. This is localized, not on the carbon atoms adjacent to the nitrogen, but primarily on the atoms labelled 'a' in the reformed pentagon on the opposing side of the vacancy (see Figs. 5B and 5F). These atoms share a weak reconstructed bond (1.73 Å), whose relative instability compared to pristine C-C bonds is reflected in an energy drop in the corresponding σ* states. Thus this additional peak at 288 eV corresponds to weakened C-C σ–bonds occurring through local strain, reconstruction and bond dilation. While these are typically found in the local environment of impurities, they are not directly linked to C-impurity bonding, hence the general nature of this peak is irrespective of impurity type (N, O, H, etc). This assignment is further supported by near edge X-ray absorption fine structure (NEXAFS) studies of graphene-based materials, where a peak at 288 eV was also observed in the proximity of folds *(33)*, edges *(34)* and divacancy species *(34)*, confirming that it represents a general signature of dilated C-C bonds associated with lattice damage, including edges, folds and grain boundaries. For similar reasons (asymmetric C-N bonds) we observe this peak at 288 eV for other defects, that will be discussed later, which are based on Stone-Thrower-Wales rearrangement with nitrogen substituted at different locations.

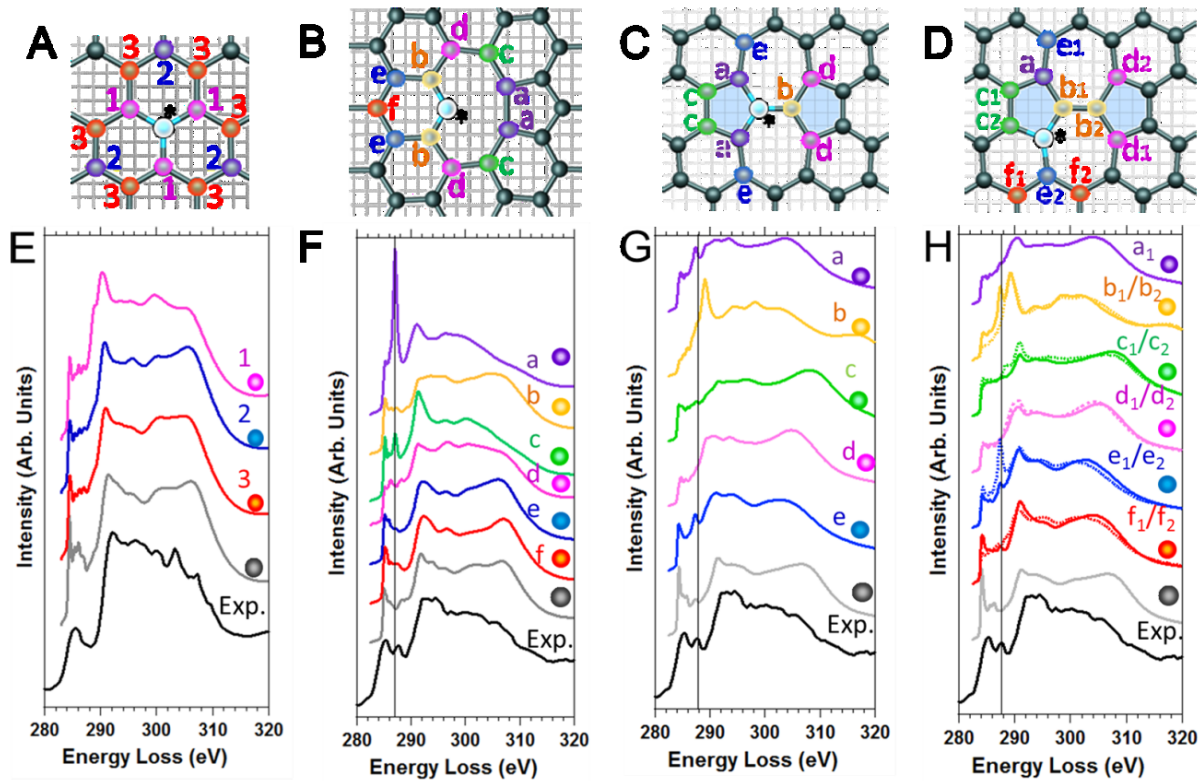

**Figure 5** *Simulated carbon 1s ionization edges for carbon atoms neighbouring nitrogen, in a (A) "graphitic" substitutional, (B) "pyridinic" nitrogen-vacancy, (C+D) "pyrrolic" substitutional with a Stone-Thrower-Wales defect configuration. (E-H) Calculated C1s spectra for the different C atoms marked in the models of Figs. 5A-D. *Nitrogen atom marked in white. Experimental spectra from Fig. 3D((i)+(ii)) and Fig. 4D((i)+(ii)).*

Figure 4C shows the N1s edge ELNES of the selected spectra (i) and (ii), along with their sum (i) + (ii). The spectral form is quite different to that of substitutional nitrogen, with a dual peak split by ~ 3 eV. The instability of this sample under the electron beam precludes the possibility of using imagery to identify the defect configuration, we rely on the spectroscopy in this case. The absence of a 288 eV C1s peak in the majority of the SI indicates that the nitrogen is not sitting in a largely amorphous carbon layer structure (*13*) but is present in a local defect configuration within an otherwise largely pristine lattice. Just as the additional 288 eV peak was assigned to a local distortion in the σ* anti-bonding states for carbon, the peak splitting in the N1s signal can also be explained via asymmetric C-N bonding. This is because the peak at approximately the same energy for substitutional nitrogen has been identified above with σ* rather than π* states. Figure 5C shows a Stone-Thrower-Wales defect, with nitrogen substituted at the defect edge. The nitrogen atom has asymmetric C-N bonds (1.357/1.410/1.448Å as compared to 1.396Å for substitutional nitrogen). In the associated simulated N1s spectrum, Figure 4D(iii), peak splitting in the N1s signal is clearly visible and corresponds well with the experimental data. Similarly to the carbon, the peak splitting is reflected primarily in the σ* states (Figure 4D(ii)). We note that the 288 eV C1s peak is also observable in some neighbouring pixels in the SI that contain no N1s signal. This is consistent with the spatially extended variation in C-C bonds associated with the Stone-Thrower-Wales defect. This allows us to assign the N1s peak at 400eV, commonly denoted "pyrrolic" nitrogen, to substitutional nitrogen in a locally asymmetric bonding configuration. The Stone-Thrower-Wales defect is an example but there are clearly other point defect structures that could contain nitrogen with asymmetric bonding, such as the 555-777 divacancy (various other defect structures are given in Supplementary Materials, only those with substitutional nitrogen in an asymmetric bonding configuration show this peak splitting). We note that this result also suggests that this peak at 400eV should always occur simultaneously with a higher energy peak, which in bulk measurements would be difficult to distinguish from substitutional nitrogen.

This study shows that, despite the technical challenges associated with motional and structural instability under the electron beam, we have resolved two types of individual nitrogen defects in single-walled carbon nanotubes, and determined their local chemical bonding environment using a combination of HAADF, EELS and DFT simulations. This is the first time that such a combination of experimental and simulated EELS spectra have been obtained at this resolution for nitrogen dopants in SWNTs, demonstrating significant differences to earlier bulk studies.

Substitutional (graphitic) nitrogen was unambiguously identified thanks to its relative stability under the 60 keV electron beam. The EELS spectroscopic signature shows a very weak $\pi^*$ peak, a sharp "molecular-like" $\sigma^*$ peak which is commonly misinterpreted as a $\pi^*$ peak, and then almost no $\sigma^*$ tail. Since bulk EELS N1s signals previously reported in the literature show a large $\sigma^*$ tail, it seems therefore likely that these are the signature of the transformation under the electron beam of substitutional nitrogen into two-fold coordinated "pyridinic" nitrogen when larger acquisition time and higher electron primary energies are used.

Analysis of a second nitrogen defect reveals that asymmetry in local sigma bonding results in peak splitting in C1s and N1s signals. Peaks at 399.8 eV and 288.0 eV in the N1s and C1s signals respectively are attributed to this effect, resolving previous literature ambiguities as to the structural origin of these peaks. This is the source of "pyrrolic" nitrogen; an example is substitutional nitrogen in the presence of a Stone-Thrower-Wales bond rotation. The 288 eV C1s peak is explained as a $\sigma^*$ signal from dilated or reconstructed C-C bonds at damage sites in the lattice, for example near to certain (but not all) impurity sites. This information could only be gained from such an atomic-resolution study.

## Acknowledgements

R.A. acknowledges funding from grants 165-119 and 165-120 from U. de Zaragoza and from ARAID foundation. We acknowledge Marcel Tencé from the LPS (CNRS-U. Paris XI, Orsay (France)) for his support with some of the EELS data analyses. We thank Shaïma Enouz-Védrenne and Jean Lou Cochon from ONERA (Châtillon and Palaiseau, France) for sample synthesis. We acknowledge the useful comments of Mike Walls from the LPS (CNRS-U. Paris XI, Orsay (France)). Part of the microscopy work was conducted in the Laboratorio de Microscopias Avanzadas at the Instituto de Nanociencia de Aragon - Universidad de Zaragoza (Spain). The research leading to these results has received funding from the European Union

Seventh Framework Program under Grant Agreement 312483 - ESTEEM2 (Integrated Infrastructure Initiative – I3) and from the French CNRS (FR3507) and CEA METSA network. C.P.E. acknowledges the NANOSIM-GRAPHENE project ANR-09-NANO-016-01 funded by the French ANR for funding. COST network MP0901 NanoTP, is acknowledged. C.P.E and X.R. thank the IMN (Nantes) and the CCIPL (Centre de Calcul Intensif des Pays de la Loire) for computing facilities.

## Author contributions

R.A., M.K. and O.S conceived the study. M.K., R.A., K.M. and O.S. performed the STEM-EELS measurements. C.P.E. and X.R. carried out DFT calculations. R.A. performed the HRTEM work. A.L. supervised nanotube synthesis. R.A., C.P.E., X.R. and O.S. wrote the paper. All authors discussed the results and edited the paper.

## Additional information

Supplementary information is available. Correspondence and requests for materials should be addressed to R.A.

## Competing financial interests

The authors declare no competing financial interests.

# Supporting Information

**Atomic Configuration of Nitrogen Doped Single-Walled Carbon Nanotubes**

Raul Arenal[1,2]*, Katia March[3], Chris P. Ewels[4], Xavier Rocquefelte[4], Mathieu Kociak[3], Annick Loiseau[5], Odile Stéphan[3]

[1]Laboratorio de Microscopias Avanzadas (LMA), Instituto de Nanociencia de Aragon (INA), Universidad de Zaragoza, Calle Mariano Esquillor, 50018 Zaragoza, Spain.

[2]ARAID Fundation, Calle Mariano de Luna, 50018 Zaragoza, Spain.

[3]Laboratoire de Physique des Solides (LPS), CNRS UMR 8502, Université Paris Sud XI, Bâtiment 510, 91405 Orsay, France.

[4] Institut des Matériaux Jean Rouxel (IMN), CNRS UMR6502, Université de Nantes, 44322 Nantes, France

[5]Laboratoire d'Etude des Microstructures (LEM), UMR 104 CNRS-ONERA, 29 Avenue de la Division Leclerc, 92322 Chatillon, France.

*To whom correspondence should be addressed: E-mail: arenal@unizar.es

*Description of the synthesis of the samples and their preparation for TEM*

The $CN_x$-SWNT sample was synthesized by laser vaporization of a graphite target mixed with Ni/Y catalyst powders in a 300 mbar $N_2$ atmosphere (*S1*). The raw powder was dispersed ultrasonically in ethanol and placed onto a copper grid coated with carbon film for the TEM-STEM studies.

*TEM and STEM instruments*

High-resolution transmission electron microscopy was performed using an imaging-side aberration-corrected FEI Titan-Cube microscope working at 80 kV, equipped with a $C_s$ corrector (CESCOR from CEOS GmbH).

The STEM-EELS-experiments were performed in a NION UltraSTEM 200 dedicated scanning transmission electron microscope, equipped with a cold field emission gun and operated at 60 kV in order to minimize knock-on damage to the sample (*S2-S3*). The energy resolution was about 0.7 eV. The collection semi-angle of the spectrometer was 50 mrad. The spectroscopic information was obtained using the spectrum-imaging acquisition mode (*S4-S5*). In order to filter the noise in the experimental data, the background-corrected EEL spectra showed in Fig. 3E, 3F, 4B, 4C, and 5 were smoothed using a Savitzky–Golay filter (second-order polynomial). To avoid contamination, the sample was heated under vacuum before being introduced into the microscope. Furthermore, the clean vacuum conditions at the sample's position in the column of the STEM (~ $1.4 \times 10^{-8}$ Torr) minimize carbon contamination. The beam current was 75 pA and the probe size was around 1.1 Å. Three different STEM imaging detectors were employed: the high angle annular dark field (HAADF) detector (with inner and outer radii of 85 and 195 mrad, respectively), the medium angle annular dark field (MAADF) detector (with inner and outer radii of 55 and 195 mrad, respectively) and the bright field (BF) detector.

## Additional TEM Studies

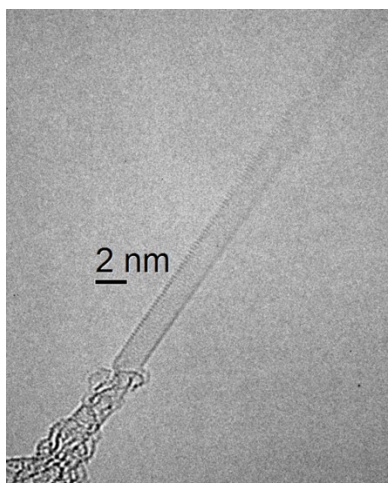

*Figure S1.* High-resolution TEM micrograph of an individual nitrogen-containing single-walled carbon nanotube ($CN_x$-SWNT). This image illustrates the instability (vibration) of the NT under the electron beam which is one of the main limiting factors for atomically-resolved analyses on NTs.

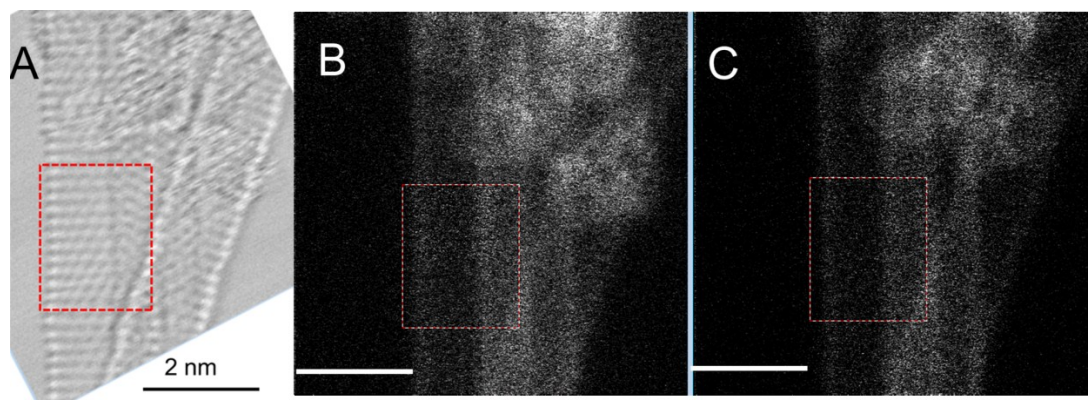

***Figure S2.*** STEM micrographs of a bundle of CN$_x$-SWNTs from which an EELS spectrum-image was collected (corresponding to Fig. 3). ***A*** and ***B*** Bright field (BF) and high angle annular dark field (HAADF) STEM images recorded before the SI acquisition. ***C*** HAADF image acquired after the SI. These images show that no apparent damage occurred during data acquisition. The scale bar of all the micrographs is 2 nm.

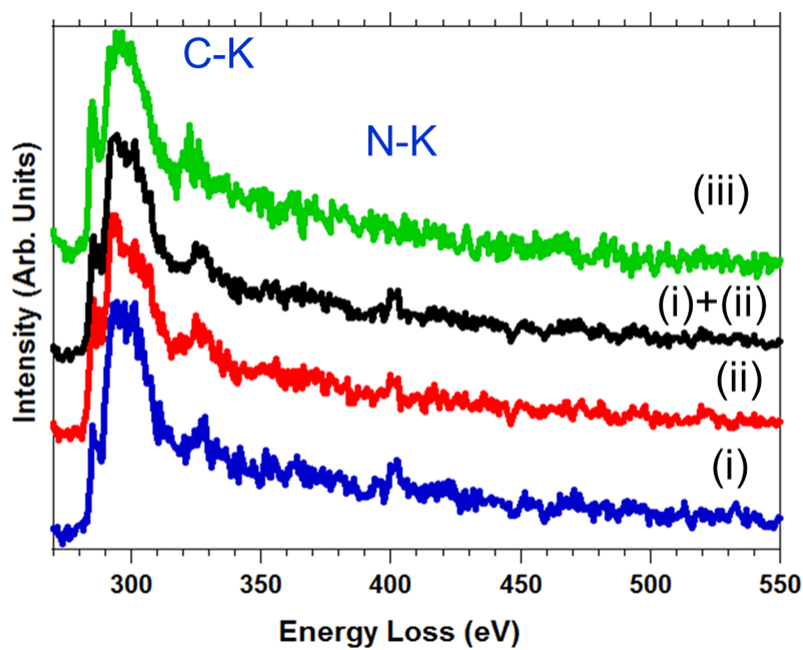

***Figure S3.*** Selection of EEL spectra from the pixels marked on Fig. 4A (red, green and the 2 pixels in blue). Each spectrum is a single pixel, except for the black one, which corresponds to the sum of the red and blue spectra. The N1s edge is observed in all these spectra, except in the green one which has been selected as a reference.

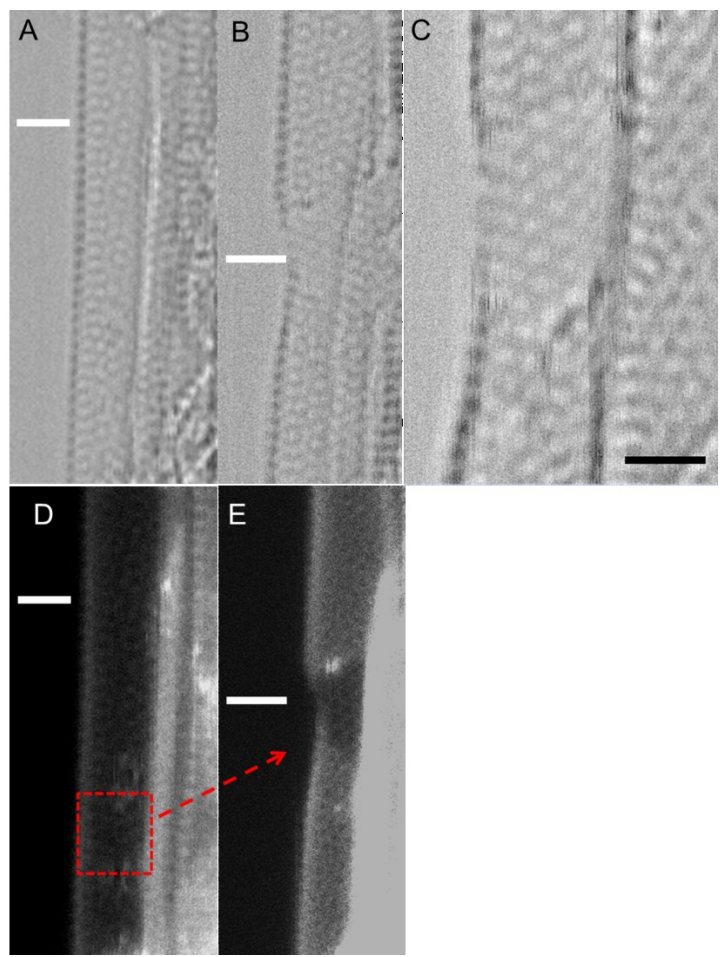

***Figure S4.*** *STEM micrographs of another bundle of $CN_x$-SWNTs from which an EELS spectrum-image was recorded (corresponding to Fig. 4). Figures **A** and **D** correspond respectively to BF and HAADF images acquired before SI collection. Figures **B, C** and **E** correspond respectively to two BF and a HAADF image acquired after SI collection. In this case, even though the acquisition conditions are the same as the previous ones (Fig. 3 and S2) the beam damage is evident. The scale bar of all the micrographs is 1 nm.*

*Irradiation effects*

We have investigated the behavior under irradiation by the electron beam of two different kinds of $CN_x$-SWNTs: those in which the defect density is low and those containing a high number of defects/distortions. From these and previous studies we suggest that the difference in the degree of crystalline quality is related to the amount of nitrogen incorporated into the hexagonal network and to the atomic (bonding) configuration of the nitrogen atoms (*S6-S7*). Thus, even though the acquisition conditions are very similar in both cases and are rather gentle, the NTs containing a higher number of defects (amount of nitrogen) are clearly much more sensitive to the electron beam, and so knock-on damage and electron-induced sputtering are more evident. This different behavior is probably related to the lower binding energies of the atoms in a distorted (highly defective) arrangement (*S8*).

## *Density Functional Theory (DFT) calculations*

The density functional theory (DFT) calculations have been carried out by using two different codes: AIMPRO (*S9 - S11*) for the geometry optimization of the structural models and WIEN2k program package (*S12*) for the calculation of the related energy loss spectra.

The structures were optimised using spin polarised density functional calculations under the local density approximation as implemented in the AIMPRO code. The charge density is fitted to plane waves with an energy cut-off of 150 Ha. Electronic level occupation was obtained using a Fermi occupation function with kT = 0.04 eV. Relativistic pseudo-potentials are generated using the Hartwingster-Goedecker-Hutter scheme (*S13*), resulting in basis sets of 22 independent Gaussian functions for carbon and 40 for nitrogen. Calculations were performed in 128 atom hexagonal supercells (a=19.543 Å, c=13.229 Å), with sufficient vacuum between layers to avoid interaction. A fine 12×12×1 k-point grid was chosen (5×5×1 for diffusion barrier calculations). Energies are converged to better than $10^{-5}$ atomic units. Atomic positions and lattice parameters were geometrically optimised until the maximum atomic position change in a given iteration dropped below $10^{-4}$ $a_0$ ($a_0$: Bohr radius). Energetic convergence with cell-size was confirmed via defect formation energies with 288 atom cells for both substitutional nitrogen and nitrogen-vacancy. The diffusion barrier for nitrogen-vacancy rearrangement was obtained using a climbing nudged-elastic-band algorithm, with 22 image structures between the nitrogen-vacancy and intermediate structure. The intermediate metastable structure was fully relaxed without constraints.

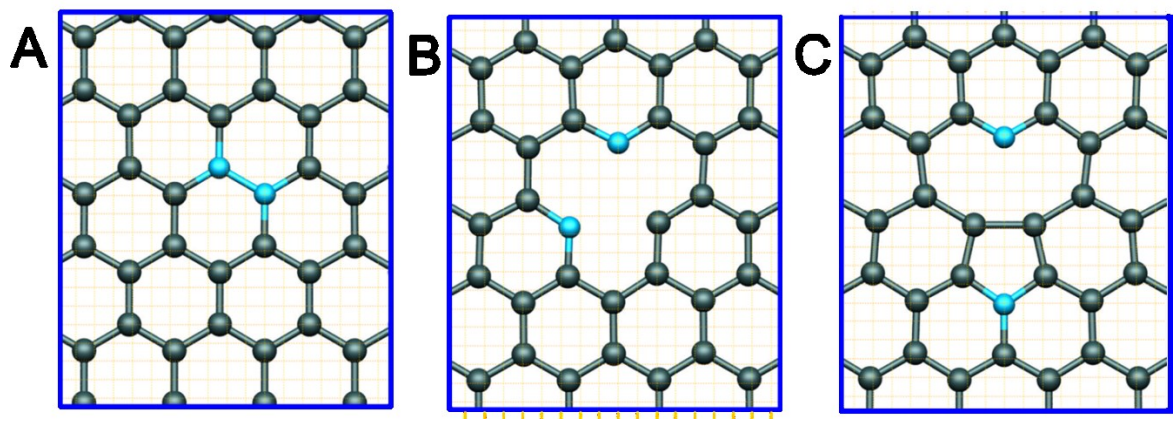

*Figure S5.* Atomic DFT optimized models of three possible nitrogen-pair configurations. Figures *A*, *B* and *D* correspond respectively to a substitutional nitrogen atom pair, two pyridinic nitrogen atoms neighboring a vacancy, and a substitutional nitrogen neighboring a nitrogen-vacancy defect.

The ELNES spectra were simulated using the Full-potential Linearized Augmented Plane Wave (FP-LAPW) method, as embodied in the WIEN2k code. The maximum l value in the expansion of the basis set inside atomic sphere was 12. The convergence of basis set is controlled by a cut-off parameter RMT×Kmax = 5, where RMT is the smallest atomic sphere radius in the unit cell and Kmax is the magnitude of the largest k vector. The self-consistency was carried out with the following radii RMT(N) = 1.27 a.u., RMT(C)=1.21 a.u., and GMAX=12 Bohr$^{-1}$.

Figure S6 shows the N 1s simulated spectra for a series of N-defect configurations. Two nitrogen positions have been considered for the Stone-Thrower-Wales defect, one with the nitrogen in the defect core shared by two heptagons and one pentagon and one with the nitrogen adjacent, shared by one heptagon, one hexagon and one pentagon. This figure provides the spectroscopic signature of each defect, allowing their identification in future atomically-resolved experiments. In addition, the experimental spectrum corresponding to Fig. 4C(i) is shown, confirming that only the simulation of nitrogen neighbouring the Stone-Thrower-Wales model gives good agreement. Indeed, the defect cannot be substitutional nitrogen, since the spectral form of the N1s differs significantly from that calculated in Figs. 3F and S6G(i). Additionally, the C1s edge in Figure 4B for the defect shows a distinct peak at 288 eV, which is not reproduced in the calculations for carbon at sites neighbouring substitutional nitrogen (see structure Fig. 5B).

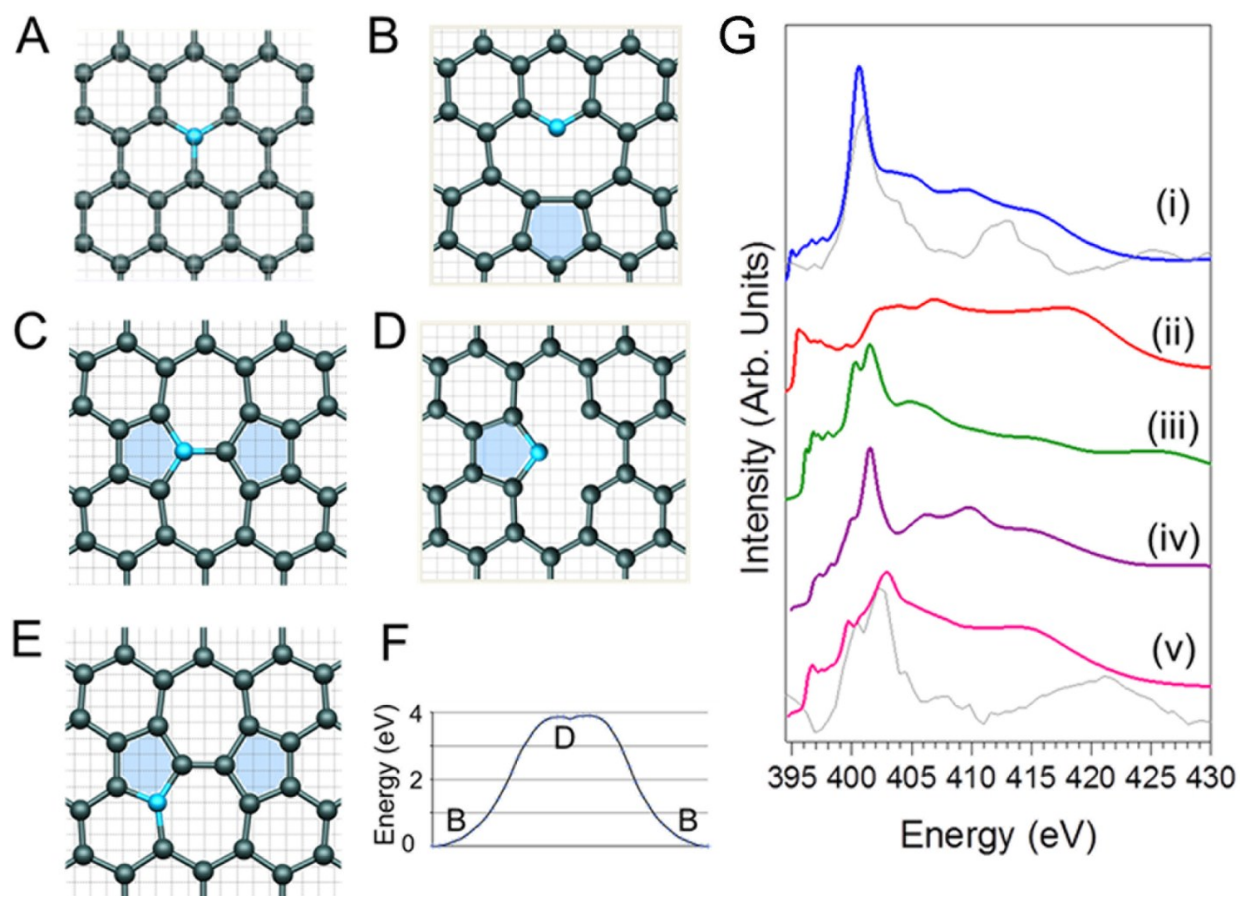

*Figure S6.* Simulated nitrogen 1s ionization edge for different nitrogen defects, in a (A) substitutional (B) nitrogen-vacancy "pyridinic", (C+E) substitutional with a Stone-Thrower-Wales defect configuration, (D) metastable transition state of B, (F) Calculated diffusion barrier for nitrogen exchange with its neighbouring vacancy in (B), via the metastable intermediate (D) at the centre of the plot. Experimental EELS spectra overlaid in grey, from Fig. 3F(iv) and 4C(i).

We also exclude the nitrogen-vacancy ("pyridinic nitrogen") centre, shown in Fig. S6B (simulated spectrum in Fig. S6G(ii)). In this case the calculated C1s edges (Fig. 5F) show the characteristic additional peak between π* and σ* for atoms on the opposite side of the vacancy forming the reconstituted pentagon bond (atoms labelled "a"). However the N1s edge does not show the experimentally observed peak splitting, and hence we can also exclude this simple model. We note that the N1s signal in this case shows a stronger π* than σ* signal, in contrast to substitutional nitrogen, since pyridinic nitrogen integrates fully into the surrounding conjugated network.

It is instructive to consider potential dynamical effects of the electron beam on nitrogen point defect structures, since peak splitting in the N1s signal could in principle occur if the defect centre changes during spectrum acquisition, resulting in overlapping spectral signatures. We note that the alternative route to obtain overlapping spectral signatures, aligned single-N atom defects in the top and bottom surfaces of the nanotube, is statistically highly improbable. Transformation from substitutional (~401 eV) to pyridinic (~399 eV) nitrogen, corresponding to the loss of a neighbouring carbon atom due to radiation damage, is very unlikely, since we are operating below the calculated knock-on threshold of carbon atoms neighbouring substitutional nitrogen (*S3*). Indeed substitutional nitrogen was seen to be beam stable in the first example presented above.

Structural rearrangement of the lattice, for example inducing bond rotations, is possible at sub-threshold irradiation (*S14*). The nitrogen-vacancy centre (Fig. S6B) can rearrange via an exchange process between the nitrogen and its neighbouring vacancy, i.e. nitrogen atom transfer from its initial C-C bond into the opposite dilated C-C bond of the vacancy. The migration pathway passes via an intermediate structure, where the N atom forms an off-centre "split vacancy" configuration, shown in Figure S6D. This structure represents the transition state in carbon vacancy migration (*S15*). With nitrogen present it is a barely metastable minimum (see Figure S6F), however it will have an extremely short lifetime. The calculated diffusion barrier for N-exchange with the vacancy is 3.87 eV (Fig. S6F), much lower than that of conventional C-C bond rotation (9.0 eV (*S16*)). Thus under the electron beam operating conditions we use here, pyridinic nitrogen would be expected to regularly interchange with its neighbouring vacancy. This would give it a spectral signature extended over several pixels and is further evidence that we do not observe the nitrogen-vacancy centre in the current sample.

Both LDA and GGA (PBE) functional have been used to estimate the impact of the functional on the simulations. In addition, we have also considered the influence of a fraction of core-hole (partially screened). Figure S7 illustrates for the substitutional nitrogen model the influence of both functional and core-hole. All the spectra are similar in shape, except the one deduced from a GGA calculation in which we do not include a core-hole. In this case, the electronic transitions start directly at the Fermi level demonstrating that valence and conduction bands are overlapping. In contrast, in both LDA and GGA with core-hole, the electronic transitions start at higher energies, as expected. Interestingly, the inclusion of a fraction of core-hole appears to be a manner to correct the deficiencies of GGA, while LDA without core-hole is already close to the experimental situation (see Fig. 3). The impact of a fraction of core-hole has been also checked for LDA and appears to be less pronounced than for GGA as evidenced in Fig. S7. It mainly leads to sharper peaks and a shift towards the lower energies of 0.6 eV. For that reason the experimental results will be compared to our simulated ELNES data deduced from LDA calculations without core-hole.

It is important to note that we are able to make direct quantitative comparison of absolute peak positions between theory and experiment, as well as comparison between relative peak positions and intensities, due to our double-reference approach. We first align all theoretical spectra with respect to the C(1s) energy of the substitutional nitrogen model. For that purpose we have considered the C(1s) energies of all models for a carbon atom far from the defect, i.e. a pristine "graphene-like" carbon. In a second step we have aligned the N(1s) simulated spectrum for substitutional nitrogen with the experimental one, leading to an energy shift of 393.6eV. This shift has then been used for all the other models.

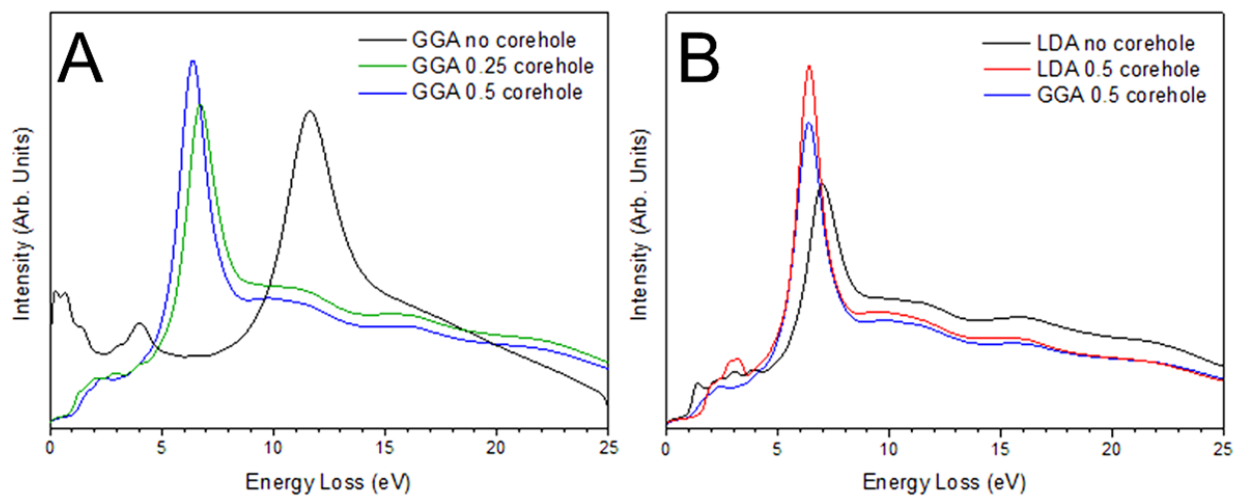

*Figure S7.* *Illustration of the influence of both functional and core-hole in the case of substitutional configuration. Figures **A** and **B** correspond to* LDA and GGA DFT calculations.

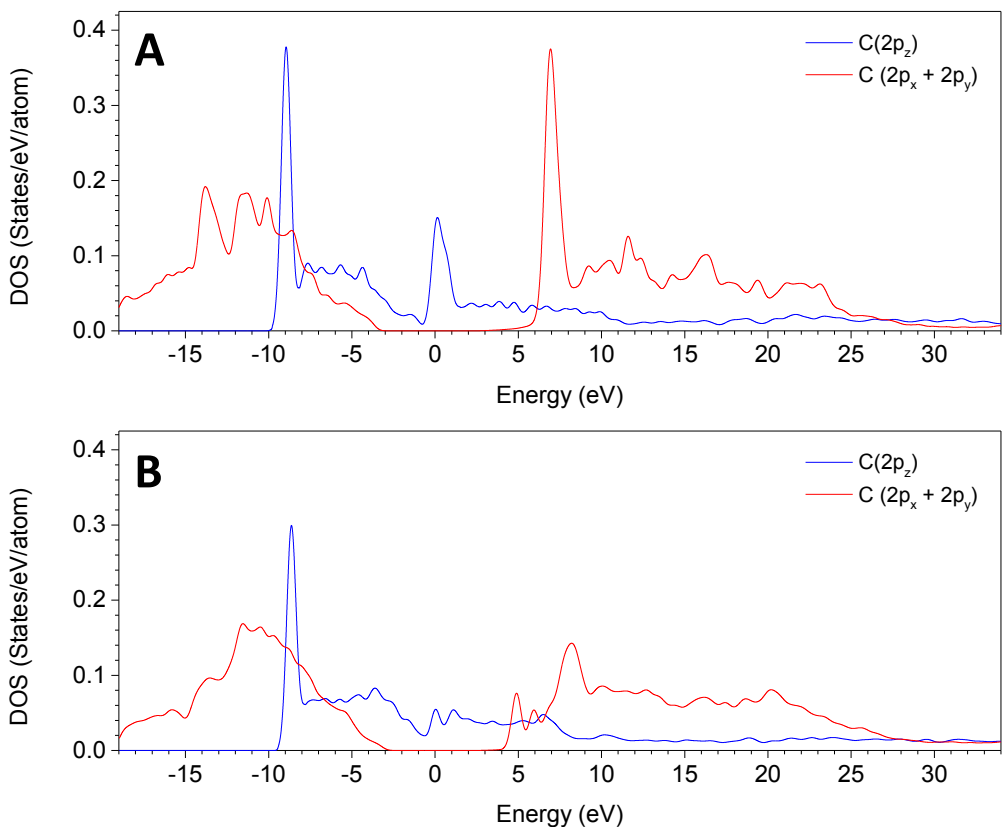

***Figure S8.*** *Partial densities of states of substitutional nitrogen (A) and substitutional nitrogen neighbouring a Stone-Thrower-Wales defect (B) models. The nitrogen 2p states are represented in order to evidence the $\pi$ and $\sigma$ interactions which are responsible of the N 1s spectra of these two models. The Fermi level is defined as the reference energy.*